\documentclass[journal,onecolumn,comsoc]{IEEEtran}
\IEEEoverridecommandlockouts
\usepackage[T1]{fontenc}
\usepackage[dvips]{psfrag}
\usepackage[dvips]{epsfig}
\usepackage{epic,color}
\usepackage[cmex10]{amsmath}
\usepackage{amsfonts}
\usepackage{amssymb}
\usepackage{amsmath}
\usepackage[cmintegrals]{newtxmath}
\usepackage{bm}
\usepackage{yfonts}
\usepackage{array}
\usepackage{bigstrut}
\usepackage{multirow}
\usepackage{dsfont}
\usepackage{cite}
\usepackage{fixltx2e}
\usepackage{accents}
\usepackage{mathtools}
\usepackage[normalem]{ulem}
\usepackage{enumerate}
\usepackage{stfloats}
\usepackage[T1]{fontenc}
\usepackage{float}
\usepackage{epstopdf}
\usepackage{mathtools}
\usepackage{algorithm}
\usepackage{algorithmicx}
\usepackage{algpseudocode}
\usepackage{algorithm,algpseudocode,float}
\usepackage{lipsum}
\usepackage{soul,xcolor}
\usepackage{xparse}
\usepackage{acronym}
\usepackage[keeplastbox]{flushend}
\usepackage{soul}

\NewDocumentCommand{\ceil}{s O{} m}{%
  \IfBooleanTF{#1} 
    {\left\lceil#3\right\rceil} 
    {#2\lceil#3#2\rceil} 
}
\NewDocumentCommand{\floor}{s O{} m}{%
  \IfBooleanTF{#1} 
    {\left\lfloor#3\right\rfloor} 
    {#2\lfloor#3#2\rfloor} 
}

\newcommand{\txt}[1]{\texttt{#1}}
\newcommand{\txtbf}[1]{\texttt{\textbf{#1}}}
\newcommand{\mbf}[1]{\mathbf{#1}}

\newcommand{\norm}[1]{\left\lVert#1\right\rVert}

\newcommand{\tblue}[1]{\textcolor{black}{#1}}

\algrenewcommand\algorithmicrequire{\textbf{Input:}}
\algrenewcommand\algorithmicensure{\textbf{Output:}}

\acrodef{5G}{fifth generation}
\acrodef{HetNet}{heterogeneous network}
\acrodef{UE}{user}
\acrodef{HD}{half-duplex}

\acrodef{FD}{full-duplex}

\acrodef{BS}{base station}
\acrodef{MBS}{macro base station}
\acrodef{SBS}{small cell base station}
\acrodef{DL}{downlink}
\acrodef{UL}{uplink}
\acrodef{CSI}{channel state information}
\acrodef{MIMO}{multiple-input multiple-output}
\acrodef{MISO}{multiple-input single-output}
\acrodef{SINR}{signal-to-interference plus noise ratio}
\acrodef{SNR}{signal-to-noise ratio}
\acrodef{AWGN}{additive white Gaussian noise}
\acrodef{MMSE}{minimum mean square error}
\acrodef{SI}{self-interference}
\acrodef{SIC}{SI cancellation}
\acrodef{CCI}{co-channel interference}
\acrodef{MUI}{multiuser interference}
\acrodef{SOC}{second-order cone}
\acrodef{SOCP}{second-order cone program}
\acrodef{RAOFDS}{resource allocation optimization for full-duplex small cell}
\acrodef{SPCA}{successive parametric convex approximation}
\acrodef{PSD}{positive semi-definite}
\acrodef{PPP}{Poisson point process}
\acrodef{DE}{decoding energy}
\acrodef{EH}{energy harvesting}
\acrodef{w.r.t.}{with respect to}
\acrodef{SE}{spectral efficiency}
\acrodef{EE}{energy efficiency}
\acrodef{AEE}{average EE}
\acrodef{SEg}{self-energy}
\acrodef{RF}{radio-frequency}
\acrodef{SWIPT}{simultaneous wireless information and power transfer}
\acrodef{RHS}{right-hand side}
\acrodef{LHS}{left-hand side}
\acrodef{KKT}{Karush-Kuhn-Tucker}

\begin{document}

\title{Is Self-Interference in Full-Duplex Communications a Foe or a Friend?}
\author{\noindent \begingroup\centering{Animesh Yadav,~\IEEEmembership{Member,~IEEE}, Octavia A. Dobre, \IEEEmembership{Senior Member, IEEE}, and H. Vincent Poor, \IEEEmembership{Fellow, IEEE}}\endgroup%
\thanks{Copyright (c) 2017 IEEE. Personal use of this material is permitted. However, permission to use this material for any other purposes must be obtained from the IEEE by sending a request to pubs-permissions@ieee.org.

Animesh Yadav and Octavia A. Dobre are with Memorial University, St. John's, NL, A1B 3X9, Canada. (e-mail: \{animeshy, odobre\}@mun.ca).

H. Vincent Poor is with Princeton University, Princeton, NJ 08544 USA (e-mail:poor@princeton.edu).}
}


\vspace{-0.2in}
\maketitle
\begin{abstract}
This paper studies the potential of harvesting energy from the self-interference of a full-duplex base station. The base station is equipped with a self-interference cancellation switch, which is turned-off for a fraction of the transmission period for harvesting the energy from the  self-interference that arises due to the downlink transmission. For the remaining transmission period, the switch is on such that the uplink transmission takes place simultaneously with the downlink transmission. A novel energy-efficiency maximization problem is formulated for  the joint design of downlink beamformers, uplink power allocations and transmission time-splitting factor. The optimization problem is nonconvex, and hence, a rapidly converging iterative algorithm is proposed by employing the successive convex approximation approach. Numerical simulation results show significant improvement in the energy-efficiency by allowing self-energy recycling. 
\end{abstract}

\IEEEpeerreviewmaketitle{\noindent }
\begin{IEEEkeywords}
Small cells, full-duplex communications, self-interference, self-energy harvesting, radio resource management.
\end{IEEEkeywords}
\vspace{-0.2in}
\section{Introduction}
Full-duplex (FD) transceivers can transmit and receive signals at the same time and frequency, and hence, provide improved spectral efficiency. However, the \ac{SI}, which suppresses the weak received signal of interest, limits their performance. With current \ac{SIC} techniques, small power transceivers are identified as being suitable for \acs{FD} deployment \cite{duplo-report-2013}. 

Recently, \ac{RF} signals have been investigated for \ac{SWIPT} \cite{Bi-Ho-Zhahg-cm-2015}. Commonly, the \ac{RF} signals' energy is harvested either by power- or time-splitting receivers. For the former (latter), the received signal power (transmission time) is divided into two parts: one used for information gathering and another one used for energy harvesting. Very recently, \acs{FD} has been combined with \ac{SWIPT} to boost both the spectral efficency and \ac{EE} of the system \cite{Zhong-Suraweera-Zheng-Krikidis-Zhang-tcom-2014, Nguyen-Duong-Tuan-Shin-Poor-tcom-2017}. Furthermore, the idea of \ac{SEg} recycling from the \ac{SI} is conceptualized in \cite{Zeng-Zhang-comml-2015, Hwang-Hwang-Kim-Lee-twc-2017, Maso-Lui-Lee-Quek-Cardosa-jsac-2015}. Both in \cite{Zeng-Zhang-comml-2015} and \cite{Hwang-Hwang-Kim-Lee-twc-2017}, \acs{FD} is used at the relay terminal and the time-splitting protocol is applied for the \ac{SEg} harvesting. The system throughput is maximized in \cite{Zeng-Zhang-comml-2015}, whereas the \ac{SNR} is maximized in \cite{Hwang-Hwang-Kim-Lee-twc-2017}. In \cite{Maso-Lui-Lee-Quek-Cardosa-jsac-2015}, the authors introduce a three-port circuit for recycling the \ac{SI} and show significant improvement in the \ac{EE}.

Since the \ac{SI} carries high energy, it could potentially be harvested for some fraction of a total transmission time. Inspired from this idea, in this paper, we consider the \ac{SEg} harvesting by the \ac{SI} at a \ac{SBS}. Particularly, we propose a time-splitting based two-phase protocol for \ac{SEg} harvesting at the \acs{FD} \ac{SBS}. The \ac{SBS} is equipped with an \ac{SIC} switch: when it is turned-on, the SIC is activated; otherwise, SIC is disabled. In the first phase, the \ac{SIC} switch is off and the \ac{SBS} sends the \textit{information-bearing} signal to its \ac{DL} \acp{UE}. The energy harvesting device at the \ac{SBS} harvests the \ac{SI} energy and also receives \textit{energy-bearing} signals from its \ac{UL} \acp{UE}. In the second phase, the \ac{SIC} switch is on, and no energy harvesting is possible from the residual \ac{SI} signal. In this phase, the \ac{SBS} continues to transmit the \textit{information-bearing} signal to \ac{DL} \acp{UE} and starts receiving the \textit{information-bearing} signals from the \ac{UL} \acp{UE}. We explore the optimal time-splitting factor that maximizes the \ac{EE} of the system along with the optimal beamforming and power allocation design for the \ac{DL} and \ac{UL} \acp{UE}, respectively. Simulation results show the significant \ac{EE} improvement offered by the proposed \ac{SEg} harvesting scheme.

\textit{Notation:} Bold uppercase and lowercase letters denote matrices and vectors, $(\cdot)^H$ and $(\cdot)^T$ represent the Hermitian and transpose operations, $|\cdot|$ and $||\cdot||_2$ denote the absolute value and $\ell_2$-norm, and $\text{tr}\{\cdot\}$ and $\mathbb{E}\{\cdot\}$ are used as the trace and expectation operators, respectively. $\mbf{I}_{N}$ represents the $N\times N$ identity matrix. $\txtbf{x}_{\geq y}$ represents a set where each element has value greater than $y$.

\vspace{-0.10in}
\section{System Model}
We consider an \acs{FD} \ac{SBS} equipped with $M_T$ transmit and $M_R$ receive antennas, which serves $K_{\txt{D}}$ and $K_{\txt{U}}$ single antenna \ac{DL} and \ac{UL} \acp{UE}, respectively. 
The sets of \ac{DL} and \ac{UL} \acp{UE} are denoted by $\mathcal{D}=\{1,\ldots, K_{\txt{D}}\}$ and $\mathcal{U}=\{1,\ldots, K_{\txt{U}}\}$, respectively. The \ac{SBS} is powered by a regular grid source, and is also equipped with an \ac{RF} power harvesting device and a rechargeable battery for energy storage. We assume a flat fading channel model, in which all channels remain unchanged for a time block of duration $T$ and change independently to new values in the next block. 

The transmission time is divided into phases of duration $\alpha T$ and $(1-\alpha)T$, where $\alpha \in (0, 1)$ is the time-splitting factor. In the first $\alpha T$ phase, the \ac{SBS} transmits the \textit{information-bearing} and receives the \textit{energy-bearing} signals to and from the \ac{DL} and \ac{UL} \acp{UE}, respectively. The \ac{SIC} switch in this phase is turned-off for \ac{SEg} harvesting. Then, the received signal at the \ac{DL} \ac{UE} $i$ is given by
\vspace{-0.1in}
\begin{equation}  \label{eq:Rx_signal_DL_1} 
y^{\txt{D}}_{1,i} = \displaystyle\mbf{h}^H_{i}\mbf{w}_{1,i}s^{\txt{D}}_{i} + \sum_{k\neq i}^{K_{\txt{D}}}\mbf{h}^H_{i}\mbf{w}_{1,k}s^{\txt{D}}_{k} + \sum_{j=1}^{K_{\txt{U}}}g_{j,i}\sqrt{p_{1,j}}s^{\txt{U}}_{j} + n^{\txt{D}}_{i},
 \vspace{-0.05in}
\end{equation} 
where $\mbf{w}_{1,i}\in \mathbb{C}^{M_T\times 1}$ and $s_i^{\txt{D}}$ with $\mathbb{E}\{|s_i^{\txt{D}}|^2\}=1$ are the beamforming vector and data of \ac{DL} \ac{UE} $i$, respectively; $p_{1,j}$ is the power coefficient allocated to the \ac{UL} \ac{UE} $j$ during $\alpha T$ phase; the vector $\mbf{h}_i\in \mathbb{C}^{M_T\times 1}$ and complex scalar $g_{j,i}$ denote the channel from the \ac{SBS} to \ac{DL} \ac{UE} $i$ and from the \ac{UL} \ac{UE} $j$ to the \ac{DL} \ac{UE} $i$, respectively; and $n^{\txt{D}}_{i}\sim \mathcal{CN}(0,\sigma^2_i)$ is complex \ac{AWGN}  at \ac{DL} \ac{UE} $i$, with variance $\sigma^2_i$. The signal vector at the receive antennas of the \ac{SBS} is given as 
\vspace{-0.1in}
\begin{IEEEeqnarray}{lll}
\mbf{r} &= \sum_{i=1}^{K_{\txt{D}}}\mbf{H}^H\mbf{w}_{1,i}s^{\txt{D}}_{i} + \sum_{j=1}^{K_{\txt{U}}}\mbf{g}_{j}\sqrt{p_{1,j}} + \mbf{n}^{r},\IEEEyesnumber
 \label{eq:Rx_nergySignal_1} 
 \vspace{-0.05in}
\end{IEEEeqnarray}  
where $\mbf{H}\in\mathbb{C}^{M_T\times M_R}$ and $\mbf{g}_{j}\in\mathbb{C}^{M_R\times 1}$ are the \ac{SI} channel matrix of the \ac{SBS} and channel vector of \ac{UL} \ac{UE} $j$ to the \ac{SBS}, respectively; and $\mbf{n}^{r}\sim \mathcal{CN}(\mbf{0}, \sigma^2_r\mbf{I}_{M_R})$ is a complex \ac{AWGN} vector at the receiver of the \ac{SBS}. Consequently, the total \ac{SEg} harvested at the \ac{SBS} in the first phase is given by
\begin{IEEEeqnarray}{lll}
E_{\text{H}} &= \eta\alpha T\mathbb{E}\{|\mbf{r}|^2\} = \eta\alpha T\Big(\sum_{i=1}^{K_{\txt{D}}}||\mbf{H}^H\mbf{w}_{1,i}||^2 + \sum_{j=1}^{K_{\txt{U}}}p_{1,j}|\mbf{g}_{j}|^2\Big),\quad \IEEEyesnumber
\label{eq:SelfEnergy_SBS} 
\vspace{-0.05in}
\end{IEEEeqnarray}
where $\eta\leq 1$ represents the energy conversion efficiency of the harvester. In \eqref{eq:SelfEnergy_SBS}, the noise term is ignored as its contribution is negligible. In the second phase, the \ac{SBS} turns the \ac{SIC} switch on, which brings the \ac{SI} power to the noise level. In this phase, the signals received at the \ac{DL} \ac{UE} $i$ and \ac{SBS} are, respectively, given as
\vspace{-0.1in}
\begin{IEEEeqnarray}{lll} 
y^{\txt{D}}_{2,i} = \displaystyle\mbf{h}^H_{i}\mbf{w}_{2,i}s^{\txt{D}}_{i} + \sum_{k\neq i}^{K_{\txt{D}}}\mbf{h}^H_{i}\mbf{w}_{2,k}s^{\txt{D}}_{k} + \sum_{j=1}^{K_{\txt{U}}}g_{j,i}\sqrt{p_{2,j}}s^{\txt{U}}_{j} + n^{\txt{D}}_{i}, \label{eq:Rx_signal_DL_2} \IEEEyesnumber\\
\mbf{y}^{\txt{U}}_{j} = \mbf{g}_{j}\sqrt{p_{2,j}}s^{\txt{U}}_{j} + \sum_{l\neq j}^{K_{\txt{U}}}\mbf{g}_{l}\sqrt{p_{2,l}}s^{\txt{U}}_{l} + \sum_{i=1}^{K_{\txt{D}}}\mbf{H}^H_{\text{on}}\mbf{w}_{2,i}s^{\txt{D}}_{i} + \mbf{n}^{\txt{U}}_j, \label{eq:Rx_signal_UL_2} 
\vspace{-0.05in}
\end{IEEEeqnarray}
where $p_{2,j}$ and $\mbf{n}^{\txt{U}}_j\sim \mathcal{CN}(\mbf{0}, \sigma^2_j\mbf{I}_{M_R})$ are the power coefficient allocated to \ac{UL} \ac{UE} $j$ and the complex \ac{AWGN} vector; and $\mbf{H}_{\text{on}}$ is the \ac{SI} channel matrix, which captures the effect of \ac{SIC}.

Using \eqref{eq:Rx_signal_DL_1} and \eqref{eq:Rx_signal_DL_2}, the received \acp{SINR} at \ac{DL} \ac{UE} $i$ in the first and second phases can be written as 
\vspace{-0.05in}
\begin{IEEEeqnarray}{lll}
&\gamma^{\txt{D}}_{l,i} =  \cfrac{|\mbf{h}^H_{i}\mbf{w}_{l,i}|^2}{\sigma^2_{i} + \sum_{k\neq i}^{K_{\txt{D}}}|\mbf{h}^H_{i}\mbf{w}_{l,k}|^2 +  \sum_{j=1}^{K_{\txt{U}}}p_{l,j}|g_{j,i}|^2},\IEEEyesnumber \label{eq:SINR_DL_1}
\vspace{-0.05in}
\end{IEEEeqnarray}
where $l = 1,\, 2$ represents the phase.
At the end of transmission time $T$, the achievable rate for \ac{DL} \ac{UE} $i$ is given as
$R^{\txt{D}}_i = \alpha\log_2(1+\gamma^{\txt{D}}_{1,i}) + (1-\alpha)\log_2(1+\gamma^{\txt{D}}_{2,i})$.

Next, for the \ac{UL} transmission, using \eqref{eq:Rx_signal_UL_2}, the received \ac{SINR} of \ac{UL} \ac{UE} $j$ at the \ac{SBS}, which applies the minimum-mean-squared error with successive interference cancellation receiver, is given by 
\vspace{-0.05in}
\begin{IEEEeqnarray}{lll}
&\gamma^{\txt{U}}_{j} = \displaystyle p_{2,j}\mbf{g}^H_{j}\mbf{X}_j^{-1}\mbf{g}_{j},\IEEEyesnumber
\label{eq:SINR_DL_1}
\vspace{-0.05in}
\end{IEEEeqnarray}
where $\mbf{X}_j \triangleq \sigma^2_{j}\mbf{I}_{M_R}+ \sum_{l>j}^{K_{\txt{U}}}p_{2,l}\mbf{g}_{l}\mbf{g}^H_{l} + \sum_{i=1}^{K_{\txt{D}}}\mbf{H}^H_{\text{on}}\mbf{w}_{2,i}\mbf{w}^H_{2,i}\mbf{H}_{\text{on}}$. Then, the achievable rate for \ac{UL} \ac{UE} $j$, at the end of transmission time $T$,  is 
$R^{\txt{U}}_j = (1-\alpha)\log_2(1+\gamma^{\txt{U}}_{j})$.

\textit{Energy usage model:} The combined energies consumed in the circuit and decoding operations are comparable or even dominate the actual transmit energy \cite{Cui-Goldsmith-Bahai-jsac-04}. Consequently, these energies play a significant role in representing the total power consumption. Thus, the total power consumed at the \ac{SBS} can be expressed as
\vspace{-0.05in}
\begin{equation}\label{eq:total_power_consumption_SBS} 
\begin{array}{l}
P^{\text{con}}_{b} = \sum\limits_{i=1}^{K_{\txt{D}}}\cfrac{\alpha}{\epsilon} \norm{\mbf{w}_{1,i}}^2 + (1-\alpha)\Big(\cfrac{\norm{\mbf{w}_{2,i}}^2}{\epsilon} +\sum\limits_{j=1}^{K_{\txt{U}}}P_{j}^{\text{dec}}(R^{\txt{U}}_j)\Big) + P_{b}^{\text{cir}}, \quad
\end{array} 
\vspace{-0.05in}
\end{equation} 
where $\epsilon \in (0,1)$ is the amplifier efficiency of the \ac{SBS}; $P_{b}^{\text{cir}} = M_TP_{\text{rf}} + P_{\text{st}}$ is the total circuit power consumption, in which $P_{\text{rf}}$ and $P_{\text{st}}$ correspond to the active \ac{RF} blocks, and to the cooling and power supply, respectively; and $P_{j}^{\text{dec}}$ is the power consumption for data decoding of the \ac{UL} \ac{UE} $j$, which is a function of the achievable rate of the \ac{UE}, i.e., for the UL \ac{UE} $j$, $P_{j}^{\text{dec}}(R^{\txt{U}}_{j}) = \beta_jR^{\txt{U}}_{j}$ where $\beta_j$ models the decoder efficiency, being decoder specific \cite{Cui-Goldsmith-Bahai-jsac-04, Yadav-Dobre-Ansari-access-2017}.

\textit{Energy efficiency function:} \tblue{In the context of 5G networks, \ac{EE} maximization is of paramount importance for both operators and users \cite{Zappone-Sanguinetti-Bacci-Jorswieck-Debbah-tsp-2016}. Thus, here as well our interest} is  to jointly optimize the \ac{DL} \ac{UE} beamformers, \ac{UL} \ac{UE} power coefficients and $\alpha$ such that the system \ac{EE} is maximized. \tblue{Unlike \cite{Zappone-Sanguinetti-Bacci-Jorswieck-Debbah-tsp-2016}, we} introduce a novel \ac{EE} function that measures the efficiency of the aggregated energies draw from the grid source at both \ac{SBS} and \ac{UL} \acp{UE} as $\eta(\mbf{w}, \mbf{p},\alpha) \triangleq  f(\mbf{w},\mbf{p},\alpha)/g(\mbf{w},\mbf{p},\alpha)$, where $f(\mbf{w},\mbf{p},\alpha)$ and $g(\mbf{w},\mbf{p},\alpha)$ capture the throughput and total grid energy consumed by the system, respectively. The sets $\txtbf{w}$ and $\txtbf{p}$ collect the optimization variables $\mbf{w}_{1,i}$ and  $\mbf{w}_{2,i}\, \forall i \in\mathcal{D}$, and $\mbf{p}_{1,j}$ and $\mbf{p}_{2,j}, \forall j\in\mathcal{U}$, respectively. In particular, the throughput is given as $f(\mbf{w},\mbf{p},\alpha) = \sum_{i=1}^{K_{\txt{D}}}R_i^{\txt{D}} + \sum_{j=1}^{K_{\txt{U}}}R_j^{\txt{U}}$, and total grid power consumption is given as $g(\mbf{w},\mbf{p},\alpha) = P_{b} + P_{u}$.
$P_{b} \triangleq  \sum_{i=1}^{K_{\txt{D}}}\alpha ||\mbf{w}_{1,i}||^2/\epsilon + \alpha P_{b}^{\text{cir}} + (1-\alpha)P_{2,b}$ denotes the consumption during the $T$ period, with $P_{2,b}$ denoting the grid power consumed during the $(1-\alpha)T$ phase. $P_{u} \triangleq \sum_{j=1}^{K_{\txt{U}}}\alpha p_{1,j} + (1-\alpha)p_{2,j}$ denotes the energy drawn from the battery source at the \ac{UL} \acp{UE} during both the $\alpha T$ and $(1-\alpha) T$ phases.
\vspace{-0.1in}
\section{Energy Efficiency Maximization}
Using the notation introduced above, the constrained \ac{EE} maximization problem is formulated as 
\vspace{-0.1in}
\begin{IEEEeqnarray*}{lcl}\label{eq:EE_Problem1}
&\underset{\begin{subarray}{c}\txtbf{w},\txtbf{p},\\ P_{2,b},\alpha \end{subarray}}{\text{max}}\, & \eta(\mbf{w},\mbf{p},\alpha) \IEEEyesnumber
\IEEEyessubnumber* \label{eq:problem_orig}\\
&\text{s.t.} & R_j^{\txt{U}} \geq \bar{r}^{\txt{U}}_j\quad \forall j\in \mathcal{U}, \label{eq:UL_UE_rate_constr}\\
&&\sum\limits_{i=1}^{K_{\txt{D}}}\alpha\norm{\mbf{w}_{1,i}}^2 +(1-\alpha)\norm{\mbf{w}_{2,i}}^2  \leq \bar{\text{P}}_{b}, \label{eq:SBS_power_constr}\\
&&\alpha p_{1,j} + (1-\alpha)p_{2,j} \leq \bar{\text{P}}_{u}\quad j\in \mathcal{U}, \label{eq:UE_power_constr}\\
&&(1-\alpha)\Big(P_{b}^{\text{cir}} + \sum_{j=1}^{K_{\txt{U}}}\beta_j\log_2(1+\gamma^{\txt{U}}_{j})+ \sum_{i=1}^{K_{\txt{D}}}\cfrac{\norm{\mbf{w}_{2,i}}^2}{\epsilon}\Big)  \leq P_{\text{H}}+ (1-\alpha)P_{2,b},\label{eq:SecondPhasePowerCon}\\
&& 0 < \alpha < 1,\label{eq:alpha_constraint}
\vspace{-0.05in}
\end{IEEEeqnarray*} 
where $P_{\text{H}}=E_{\text{H}}/T$, and $\bar{\text{P}}_{b}$ and $\bar{\text{P}}_{u}$ denote the maximum transmit powers of the \ac{SBS} and \ac{UL} \ac{UE}, respectively. Constraint \eqref{eq:UL_UE_rate_constr} ensures that each \ac{UL} \ac{UE} achieves the minimum specified data-rate of $\bar{r}^{\txt{U}}_j$. Constraints \eqref{eq:SBS_power_constr} and \eqref{eq:UE_power_constr} represent the restrictions on the maximum transmit powers of the \ac{SBS} and the \ac{UL} \acp{UE}, respectively. Constraint \eqref{eq:SecondPhasePowerCon} restricts the \ac{SBS} to use the harvested \ac{SEg} in the second phase, if it is sufficient; otherwise, the \ac{SBS} draws the energy from the grid source to sustain the transmissions. Evidently, \eqref{eq:EE_Problem1} is a nonconvex problem and obtaining an optimal solution is challenging and converges slowly. Hence, we seek a rapidly converging suboptimal solution in the following section. 
\vspace{-0.1in}
\section{Proposed Solution Method}
There are two main steps involved to arrive at the rapidly converging solution. In the \textit{first step}, we perform a few equivalent transformations on \eqref{eq:EE_Problem1}, similar to \cite{Yadav-Dobre-Ansari-access-2017}  and \cite{Nguyen-Tran-Pirinen-Latva-aho-twc-2014}, to expose the hidden convexity and gain tractability. Accordingly, the resulting problem is expressed equivalently as 
\vspace{-0.1in}
\begin{IEEEeqnarray*}{lcl}\label{eq:Problem1_equivalent}
&\underset{\Xi}{\text{max}} & \quad q^2 \qquad \qquad \qquad\qquad \qquad \qquad  \IEEEyesnumber
\IEEEyessubnumber* \label{eq:obj_equivalent}\\
&\text{s.t.} & z^{\txt{U}}_j \geq \bar{r}^{\txt{U}}_j\, \quad \forall j\in \mathcal{U}, \label{eq:UL_UE_rate_constr_equivalent}\\
&& \big(\mbf{1}^T\mbf{z}^{\txt{D}}_{1} + \mbf{1}^T\mbf{z}^{\txt{D}}_{2} + \mbf{1}^T\mbf{z}^{\txt{U}}\big)\tau \geq q^2 \label{eq:obj_const_equivalent} \\
&& \cfrac{1}{c} \geq  \sum_{i=1}^{K_{\txt{D}}}\cfrac{\norm{\mbf{w}_{1,i}}^2}{\epsilon} + P_{b}^{\text{cir}}+ \cfrac{p_{2,b}}{\alpha}-p_{2,b}   + \sum_{j=1}^{K_{\txt{U}}}\Big(p_{1,j}  + \cfrac{p_{2,j}}{\alpha} - p_{2,j}\Big) \label{eq:power_used_const_equivalent}\\
&& c \geq \alpha\tau \label{eq:power_used_const_equivalent_1}\\
&& \frac{|\mbf{h}^H_{i}\mbf{w}_{1,i}|^2}{u^{\txt{D}}_{1,i}} \geq \sigma^2_{i} + \sum_{k\neq i}^{K_{\txt{D}}}|\mbf{h}^H_{i}\mbf{w}_{1,k}|^2 +  \sum_{j=1}^{K_{\txt{U}}}p_{1,j}|g_{j,i}|^2\, \forall i \in  \mathcal{D}, \nonumber\\*\label{eq:ph1_const1_equivalent}\\
&&\mbf{h}^H_{i}\mbf{W}_{2,i}\mbf{h}_{i} \geq u^{\txt{D}}_{2,i}b_{i} \, \forall i \in  \mathcal{D}, \label{eq:ph2_const1_equivalent}\\
&& u^{\txt{D}}_{1,i} + 1 \geq (t^{\txt{D}}_{1,i})^{1/\alpha}, \, u^{\txt{D}}_{2,i} + 1 \geq (t^{\txt{D}}_{2,i})^{1/(1-\alpha)} \, \forall i \in  \mathcal{D}, \label{eq:ph1_and_ph2_const2_equivalent}\\
&& b_{i} \geq \displaystyle \sigma^2_{i} + \sum_{k\neq i}^{K_{\txt{D}}}\mbf{h}^H_{i}\mbf{W}_{2,k}\mbf{h}_{i} + \sum_{j=1}^{K_{\txt{U}}}p_{2,j}|g_{j,i}|^2  \, \forall i \in  \mathcal{D}, \label{eq:ph2_const2_equivalent}\\
&& x^2_{j}\mbf{g}^H_{j}\mbf{X}^{-1}\mbf{g}_{j} \geq u^{\txt{U}}_{j} \quad \forall j \in  \mathcal{U},\label{eq:ph2_const3_equivalent}\\
&& u^{\txt{U}}_{j} + 1 \geq (t^{\txt{U}}_{j})^{1/(1-\alpha)} \, \forall j \in  \mathcal{U}, \label{eq:ph2_const4_equivalent}\\
&& p_{2,j}\geq x^2_{j} \, \forall j \in  \mathcal{U}, \label{eq:ph2_const5_equivalent}\\
&& t^{\txt{D}}_{1,i} \geq e^{z^{\txt{D}}_{1,i}},\, t^{\txt{D}}_{2,i} \geq e^{z^{\txt{D}}_{2,i}}, \, t^{\txt{U}}_{j} \geq e^{z^{\txt{U}}_{j}} \, \forall (i,j) \in  (\mathcal{D}, \mathcal{U}), \label{eq:ph1_and_ph2_const3_equivalent}\\
&& \eta \bar{P}_{H} \geq \cfrac{P_{b}^{\text{cir}}}{\alpha} - P_{b}^{\text{cir}} + \sum_{j=1}^{K_{\txt{U}}}\beta_j\big(\cfrac{z^{\txt{U}}_{j}}{\alpha} - z^{\txt{U}}_{j}\big) - \cfrac{p_{2,b}}{\alpha} + p_{2,b}  + \cfrac{1}{\epsilon}\sum_{i=1}^{K_{\txt{D}}}\cfrac{\text{tr}(\mbf{W}_{2,i})}{\alpha}-\text{tr}(\mbf{W}_{2,i}),\label{eq:power_EH_const_equivalent}
\end{IEEEeqnarray*} 
\begin{IEEEeqnarray*}{lcl}
&& \cfrac{\bar{P}_b}{\alpha}  \geq \sum\limits_{i=1}^{K_{\txt{D}}}\Big(\norm{\mbf{w}_{1,i}}^2 +\cfrac{\text{tr}(\mbf{W}_{2,i})}{\alpha}-\text{tr}(\mbf{W}_{2,i})\Big),\label{eq:BS_power_const_equivalent} \IEEEyessubnumber*\\
&&  \cfrac{\bar{P}_u}{\alpha} \geq p_{1,j} + \cfrac{p_{2,j}}{\alpha}-p_{2,j}\, \forall j\in \mathcal{U}, \qquad \qquad \qquad\qquad \qquad \qquad \qquad \qquad \label{eq:UE_power_const_equivalent}   \\
&& \eqref{eq:alpha_constraint}, \label{eq:positive_const_equivalent}\\
&& \text{rank}(\mbf{W}_{2,i}) = 1, \forall i \in \mathcal{D}, \label{eq:rank_constraint}
\end{IEEEeqnarray*}
where $\bar{P}_H \triangleq P_H/\alpha$ and $\mbf{W}_{2,i} = \mbf{w}_{2,i}\mbf{w}^H_{2,i}$ is a rank-1 \ac{PSD} matrix. The introduction of $\mbf{W}_{2,i}$ helps convexify \eqref{eq:ph2_const3_equivalent} \cite{Nguyen-Tran-Pirinen-Latva-aho-twc-2014}, which is otherwise a difficult constraint to handle. For notational compactness, a set $\Xi = \{\txtbf{w}, \txtbf{W}, \txtbf{p}, \alpha, \txtbf{z}, \txtbf{u}, \txtbf{t}, \txtbf{b}, \txtbf{x}, c, \tau, q\}$ is introduced, which collects all the optimization variables, where $\txtbf{W}_{\geq 0}, \txtbf{z}_{\geq 0}, \txtbf{u}_{\geq 0}, \txtbf{t}_{\geq 1}$ collect $\{\mbf{W}_{2,1}, \ldots, \mbf{W}_{2,K_{\txt{D}}}\}$, $\{\mbf{z}^{\txt{D}}_1, \mbf{z}^{\txt{D}}_2, \mbf{z}^{\txt{U}}\}$, $\{\mbf{u}^{\txt{D}}_1, \mbf{u}^{\txt{D}}_2, \mbf{u}^{\txt{U}}\}$ and $\{\mbf{t}^{\txt{D}}_1, \mbf{t}^{\txt{D}}_2, \mbf{t}^{\txt{U}}\}$, respectively. $\mbf{z}^{\txt{D}}_1 = [z^{\txt{D}}_{1,1},\ldots, ^{\txt{D}}_{1, K_{\txt{D}}}]^T$, $\mbf{z}^{\txt{D}}_2 = [z^{\txt{D}}_{2,1},\ldots, z^{\txt{D}}_{2, K_{\txt{D}}}]^T$, $\mbf{z}^{\txt{U}} = [z^{\txt{U}},\ldots, z^{\txt{U}}_{K_{\txt{D}}}]^T$, $\mbf{u}^{\txt{D}}_1 \in \{u^{\txt{D}}_{1,1},\ldots, u^{\txt{D}}_{1, K_{\txt{D}}}\}$, $\mbf{u}^{\txt{D}}_2 \in \{u^{\txt{D}}_{2,1},\ldots, u^{\txt{D}}_{1, K_{\txt{D}}}\}$, $\mbf{u}^{\txt{U}} \in \{u^{\txt{U}},\ldots, u^{\txt{U}}_{ K_{\txt{U}}}\}$, $\mbf{t}^{\txt{D}}_1 \in \{t^{\txt{D}}_{1,1},\ldots, t^{\txt{D}}_{1, K_{\txt{D}}}\}$, $\mbf{t}^{\txt{D}}_2 \in \{t^{\txt{D}}_{2,1},\ldots, t^{\txt{D}}_{2, K_{\txt{D}}}\}$, $\mbf{t}^{\txt{U}} \in \{t^{\txt{U}},\ldots, t^{\txt{U}}_{K_{\txt{D}}}\}$, $\txtbf{b} \in \{b_1,\ldots, b_{K_{\txt{D}}}\}$, $\txtbf{x} \in \{x_1,\ldots, x_{K_{\txt{U}}}\}$, $c\geq 0$, $\tau\geq 0$, $q\geq 0$ are slack variables. It is easy to see that a solution to \eqref{eq:Problem1_equivalent} is also feasible for \eqref{eq:EE_Problem1}. Moreover, all the constraints \eqref{eq:UL_UE_rate_constr_equivalent}-\eqref{eq:power_EH_const_equivalent} are active at optimality, and hence, \eqref{eq:Problem1_equivalent} is an equivalent formulation of \eqref{eq:EE_Problem1}. 
Note that, to write \eqref{eq:obj_const_equivalent} as a \ac{SOC} constraint, we introduce $q^2$ in the objective function; however, its maximization is a nonconvex problem. Hence, we equivalently replace the objective function with $q$, which also maximizes $q^2$. Next, to achieve further tractability, we relax the nonconvex rank-1 constraint \eqref{eq:rank_constraint} by dropping it. Now, \eqref{eq:Problem1_equivalent} can be equivalently expressed as
\vspace{-0.05in}
\begin{IEEEeqnarray*}{lcl}\label{eq:Problem1_equivalent_1}
\max_{\Xi}\{q|\eqref{eq:UL_UE_rate_constr_equivalent}-\eqref{eq:positive_const_equivalent}\}. \IEEEyesnumber
\vspace{-0.05in}
\end{IEEEeqnarray*}
In the \textit{second step}, we identify the nonconvex parts of \eqref{eq:Problem1_equivalent_1} and linearize them with a first-order Taylor approximation around the point of operation \cite{Marks-Wright-or-78}. This step leads to an iterative procedure and a local solution to \eqref{eq:Problem1_equivalent_1}. In \eqref{eq:Problem1_equivalent}, the constraints \eqref{eq:power_used_const_equivalent}-\eqref{eq:ph1_and_ph2_const2_equivalent}, \eqref{eq:ph2_const4_equivalent}, \eqref{eq:power_EH_const_equivalent}-\eqref{eq:UE_power_const_equivalent} are nonconvex. Particularly, the nonconvexity in \eqref{eq:power_used_const_equivalent}, \eqref{eq:ph1_const1_equivalent}, \eqref{eq:BS_power_const_equivalent} and \eqref{eq:UE_power_const_equivalent} is due to the convex function of form $f_1(x,y) = |x|^2/y, \forall x \in \mathbb{C}^N, y \in \mathbb{R}^{+}$, on the greater side of the inequalities. Functions of this form can be approximated, around a point $(x^{(n)}, y^{(n)})$ at the $n$th iteration, as $F^{(n)}_1(x, y) = 2\mathfrak{R}(x^{(n)}x)/y^{(n)} - |x^{(n)}|^2y^{(n)}/(y^{(n)})^2$. The constraints \eqref{eq:power_used_const_equivalent}, \eqref{eq:ph1_and_ph2_const2_equivalent}, \eqref{eq:ph2_const4_equivalent}, \eqref{eq:power_EH_const_equivalent}, \eqref{eq:BS_power_const_equivalent} and \eqref{eq:UE_power_const_equivalent} also have nonconvexity due the presence of functions of the forms  $f_2(x,y)= x/y$ and $f_3(x,y) = x^{(1/y)}, \forall x \in \mathbb{C}^N, y \in (0,1)$, which we linearize around a point $x^{(n)}, y^{(n)}$, as $F^{(n)}_{k}(x, y) = f_{k}(x^{(n)}, y^{(n)}) + \langle\nabla f_{k}(x^{(n)}, y^{(n)}), (x, y) - (x^{(n)}, y^{(n)})\rangle,\, k \in \{2,3\}$ \cite{Tuy-book-16}. The constraints \eqref{eq:power_used_const_equivalent_1} and \eqref{eq:ph2_const1_equivalent} have nonconvexity on the lesser side of the inequalities of the form $f_4(x,y)= xy, \forall x \in \mathbb{C}^N, y \in \mathbb{R}^{+}$. Using the result from \cite{Tran-Hanif-Tolli-Juntti-spl-2012}, we replace $f_4(x,y)$ with its convex upper bound around a point $(x^{(n)}, y^{(n)})$ as $F^{(n)}_4(x, y, \phi^{(n)}) = 0.5(\phi^{(n)}(x^{(n)})^2 + (y^{(n)})^2/\phi^{(n)}), \forall \phi^{(n)} = y^{(n)}/x^{(n)}> 0$. \tblue{The approximations employed above satisfy the following three conditions \cite{Marks-Wright-or-78}: i) $F^{(n)}(x)\leq f(x), \forall x$; ii) $F^{(n)}(x^{(n-1)})= f(x^{(n-1)})$; iii) $\nabla F^{(n)}(x^{(n-1)})= \nabla f(x^{(n-1)})$, and hence, the convergence of the iterative procedure is ensured.} Now, by replacing the nonconvex parts of the constraints with the approximations discussed above, \eqref{eq:Problem1_equivalent_1} can be formulated as a convex problem at the $n$th iteration as 
\vspace{-0.08in}
\begin{IEEEeqnarray*}{lcl}\label{eq:Problem1_approx}
&\underset{\Xi}{\text{max}} & \quad q \qquad \qquad \qquad\qquad \qquad \qquad\qquad  \IEEEyesnumber
\IEEEyessubnumber* \label{eq:obj_approx}\\
&\text{s.t.} & F^{(n)}_1(1,c) \geq  \sum_{i=1}^{K_{\txt{D}}}\norm{\mbf{w}_{1,i}}^2 + P_{b}^{\text{cir}}+ F^{(n)}_2(\alpha, p_{2,b}) - p_{2,b}    + \sum_{j=1}^{K_{\txt{U}}}\big(p_{1,j} + F^{(n)}_2({\alpha, p_{2,j}}\big)- p_{2,j} \label{eq:power_used_const_approx}\\
&& c \geq F^{(n)}_4(\alpha, \tau,\phi^{(n)}_1) \label{eq:power_used_const_approx_1}\\
&& F^{(n)}_1(\mbf{w}_{1,i},u^{\txt{D}}_{1,i})  \geq \sigma^2_{i} + \sum_{k\neq i}^{K_{\txt{D}}}|\mbf{h}^H_{i}\mbf{w}_{1,k}|^2 +  \sum_{j=1}^{K_{\txt{U}}}p_{1,j}|g_{j,i}|^2\,\forall i \in  \mathcal{D}, \label{eq:ph1_const1_approx}\\
&&\mbf{h}^H_{i}\mbf{W}_{2,i}\mbf{h}_{i} \geq F^{(n)}_4(u^{\txt{D}}_{2,i}, b_{i},\phi^{(n)}_{2,i}) \, \forall i \in  \mathcal{D}, \label{eq:ph2_const1_approx}\\
&& u^{\txt{D}}_{1,i} + 1 \geq F^{(n)}_3(t^{\txt{D}}_{1,i}, \alpha), \, u^{\txt{D}}_{2,i} + 1 \geq F^{(n)}_3(t^{\txt{D}}_{2,i}, \bar{\alpha}) \, \forall i \in  \mathcal{D},\qquad \label{eq:ph1_and_ph2_const2_approx}\\
&& u^{\txt{U}}_{j} + 1 \geq F^{(n)}_3(t^{\txt{U}}_{j}, \bar{\alpha}) \, \forall j \in  \mathcal{U}, \label{eq:ph2_const4_approx}\\
&& \eta \bar{P}_{H} \geq \cfrac{\bar{\alpha}P_{b}^{\text{cir}}}{\alpha} + \sum_{j=1}^{K_{\txt{U}}}\beta_j\big(F^{(n)}_2(\alpha, z^{\txt{U}}_{j}) - z^{\txt{U}}_{j}\big) - F^{(n)}_2(\alpha, p_{2,b})
 + p_{2,b} + \cfrac{1}{\epsilon}\sum_{i=1}^{K_{\txt{D}}}F^{(n)}_2(\alpha, \text{tr}(\mbf{W}_{2,i}))-\text{tr}(\mbf{W}_{2,i}),\label{eq:power_EH_const_approx}\IEEEyessubnumber* \\
&& \cfrac{\bar{P}_b}{\alpha}  \geq \sum\limits_{i=1}^{K_{\txt{D}}}\Big(\norm{\mbf{w}_{1,i}}^2 + F^{(n)}_2(\alpha, \text{tr}(\mbf{W}_{2,i})) - \text{tr}(\mbf{W}_{2,i})\Big),\label{eq:BS_power_const_approx}\\
&&  \cfrac{\bar{P}_u}{\alpha} \geq p_{1,j} + F^{(n)}_2(\alpha, p_{2,j}) - p_{2,j}\, \forall j\in \mathcal{U}, \label{eq:UE_power_const_approx} \\
&& \eqref{eq:UL_UE_rate_constr_equivalent}, \eqref{eq:obj_const_equivalent}, \eqref{eq:ph2_const2_equivalent}, \eqref{eq:ph2_const3_equivalent}, \eqref{eq:ph2_const5_equivalent}, \eqref{eq:ph1_and_ph2_const3_equivalent}, \eqref{eq:alpha_constraint}, \label{eq:positive_const_approx}
\end{IEEEeqnarray*}
where $\bar{\alpha} \triangleq 1-\alpha$. Pseudocode for solving \eqref{eq:Problem1_approx} is outlined in Algorithm~1. Note that the constraints in \eqref{eq:ph1_and_ph2_const3_equivalent} are exponential cones, which we approximate as a set of \ac{SOC} constraints \cite{Tal-Nemirovski-mor-2001, Nguyen-Yadav-Ajib-Assi-twc-2016} in Algorithm~1. Since the problem is upper bounded due to power constraints, the algorithm generates a monotonic non-decreasing sequence of objective function values and converges to a \ac{KKT} point of \eqref{eq:Problem1_equivalent_1} \cite{Marks-Wright-or-78}. The detailed proof follows similar lines as the one discussed in \cite{Nguyen-Yadav-Ajib-Assi-twc-2017}, and hence, is omitted here for brevity.

The feasible initial point to Algorithm~1 is generated by solving the following problem; 
$\max_{\Xi, \bm{\mu}\leq 0} \quad \{q + \mbf{1}^T\bm{\mu}|\eqref{eq:power_used_const_approx}-\eqref{eq:positive_const_approx}/\eqref{eq:UL_UE_rate_constr_equivalent}\}$ subject to $\bar{r}^{\txt{U}}_j + \mu_j\leq z^{\txt{U}}_j \, \forall j \in \mathcal{U}$, where $\bm{\mu}=[\mu_1, \ldots, \mu_{K_{\txt{U}}}]^T$ are the newly introduced variables. The feasible initial point is obtained when $\bm{\mu}=\mbf{0}$ and requires three iterations at most.\phantom{\cite{Sidiropoulos-Davidson-Luo-tsp-06,3gpp-36-828-report-2012}}

\begin{figure}[h!]
\centerline{\epsfig{figure=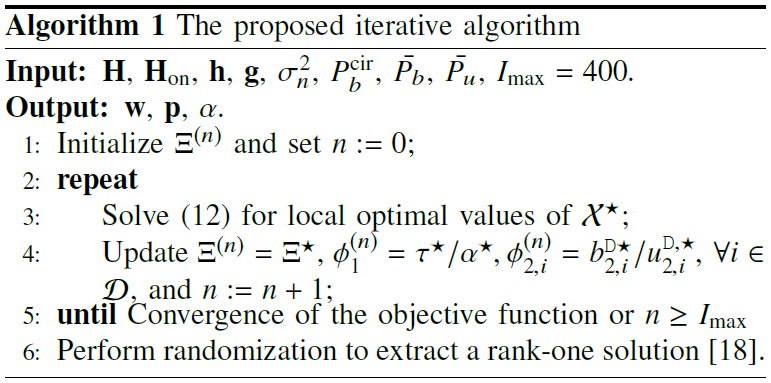,width=2.75in}}
\label{ALG:Offline_Algo-1}
\end{figure}

\begin{figure}[h!]
\centerline{\epsfig{figure=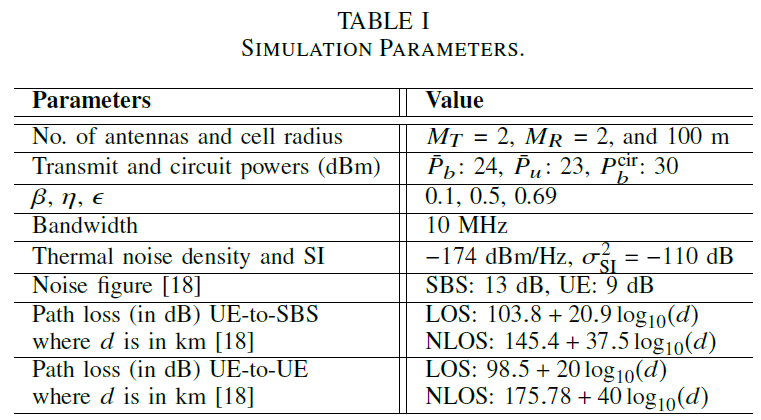,width=3.0in}}
\end{figure} \label{tab:simulation_parameters}

\section{Numerical Results}
In this section, a performance evaluation of the proposed \ac{SEg} harvesting scheme is presented. The parameters used in simulations with their values are listed in Table~I. The algorithm is implemented using the CVX parser \cite{cvx} and \txt{mosek} as an internal solver. The minimum data-rate required for each \ac{UL} \ac{UE} is set to $\bar{r}^{\txt{U}}_j = 1$ Mbit/s. A small cell of radius $100$ m is considered with $K_{\txt{D}} = K_{\txt{U}} = 2$ uniformly distributed \acp{UE} within the cell area. The channels are Rayleigh faded with each coefficient following the $\mathcal{CN}(0,1)$ distribution. The \ac{SI} channel is modeled as Rician, with Rician factor $K=1$, i.e., $\mbf{H}_{\text{on}}\sim \mathcal{CN}_{M_T\times M_R}\Big(\sqrt{\sigma^2_{\text{SI}}K/(K+1)}\bar{\mbf{H}},(\sigma^2_{\text{SI}}/(1+K))\mbf{I}_{M_R}\otimes \mbf{I}_{M_T}\Big)$, where $\bar{\mbf{H}}$ is a deterministic matrix and $\otimes$ denotes the Kronecker product. $\sigma^2_{\text{SI}}$ denotes the ratio of the average \ac{SI} powers before and after the \ac{SIC}. \tblue{Results are obtained based on} $1000$ runs.

\begin{figure}[h!]
\centerline{\epsfig{figure=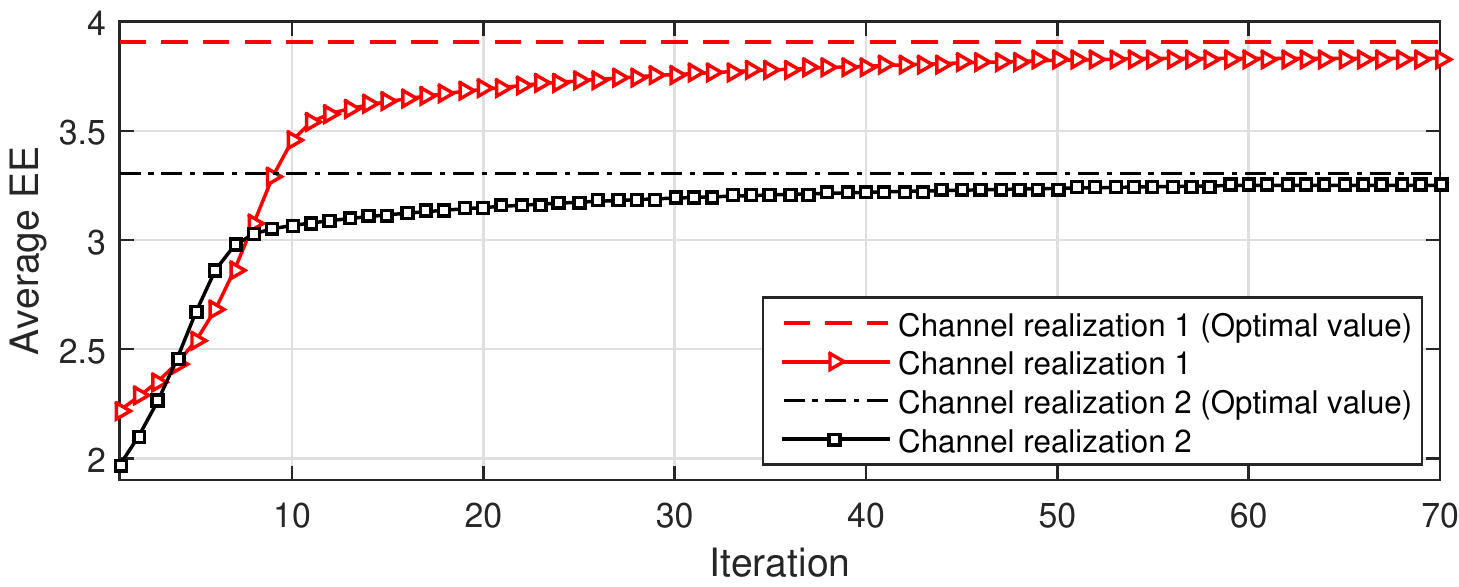,width=3.0in}}
\begin{center}
\vspace{-0.1in}
\caption{Convergence behavior of Algorithm~1 for two independent channel realizations with $\bar{P}_b=25$ dBm, \tblue{ $M_T = 1$, and $M_R = 1$}.}\label{fig:convergence_algo}
\end{center}
\end{figure}

In Fig.~\ref{fig:convergence_algo}, we show the convergence behavior of the iterative Algorithm~1. The objective function values are plotted versus the number of iterations for \tblue{two} independent random channels states and $\bar{P}_b=25$ dBm. We observe that Algorithm~1 converges in less than fifty iterations for all three channel realizations. \tblue{For benchmarking purpose, the objective function values are also compared with the global ones, which are obtained using the branch-reduce-and-bound algorithm \cite{Bjornson-Zheng-Bengtsson-Ottersten-tsp-2012}.}

\begin{figure}[h!]
\centerline{\epsfig{figure=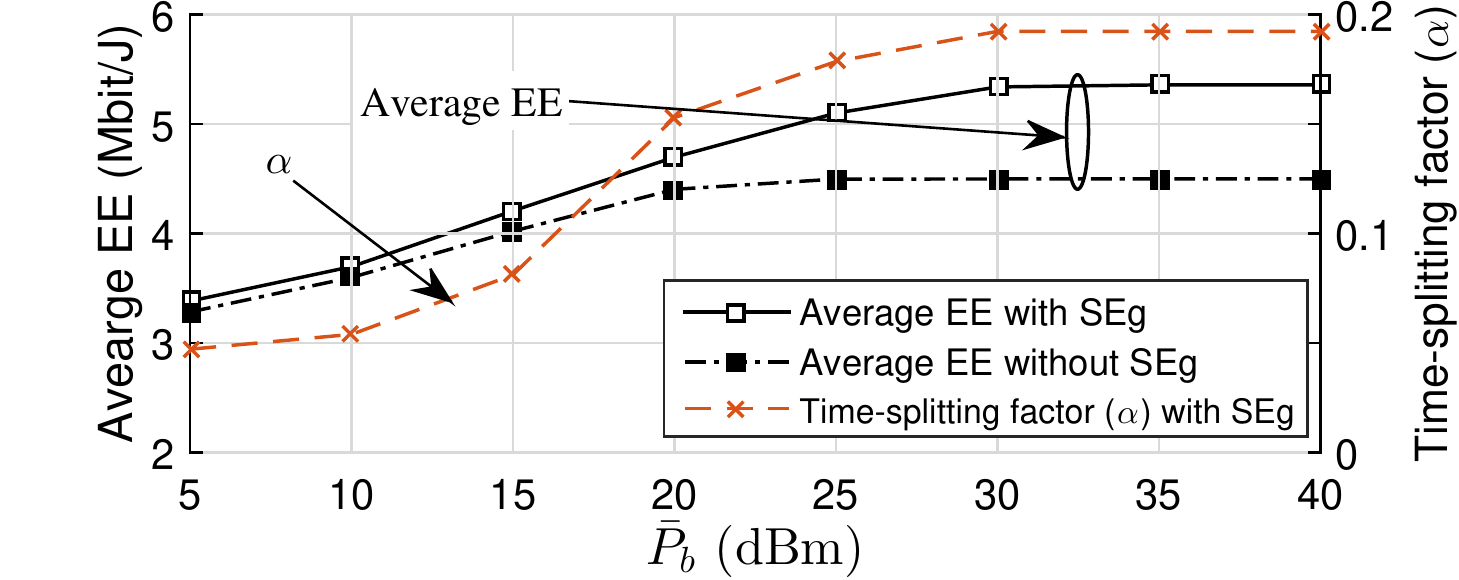,width=3.0in}}
\begin{center}
\vspace{-0.1in}
\caption{The average \ac{EE} (left-hand side y-axis) and the time-splitting factor (right-hand side y-axis) versus the maximun transmit power of the \ac{SBS}.}\label{fig:EE_vs_Pbs}
\end{center}
\end{figure}

In Fig.~\ref{fig:EE_vs_Pbs}, the \ac{AEE} with and without the \ac{SEg} harvesting scheme versus the transmit power of the \ac{SBS} is plotted. It can be seen that the average gain of the scheme that harvests \ac{SEg} is significantly higher than the one that does not harvest. The \ac{AEE} of both schemes saturates in the high $\bar{P}_b$ regime; however, the former saturates later than the latter. \tblue{In low-power regime, for the proposed scheme, the \ac{SBS} harvests less \ac{SEg} and draws more energy from the grid source for decoding the UL users data, and hence, the \ac{AEE} drops.} Note that the \ac{AEE} of the latter is obtained by using Algorithm~1 with fixed $\alpha = 0$. The time-splitting factor $\alpha$ is also shown on the right-hand side y-axis of the figure. Its values increase with $\bar{P}_b$ but saturates in the high $\bar{P}_b$ regime. 
  
Lastly, we have observed that in the first phase Algorithm~1 allocates zero power to each \ac{UL} \ac{UE} for all values of $\bar{P}_b$, and accordingly, the \ac{SBS} harvests only the \ac{SEg}.

\section{Conclusion}
We have considered a FD SBS, in which, by installing an additional on and off \ac{SIC} switch, the \acs{FD} \ac{SBS} can harvest \ac{SEg} from the \ac{SI}. The \ac{SEg} is harvested when the \ac{SIC} switch is off; otherwise, there is no harvesting. The fraction of the transmission time for which the \ac{SIC} switch is on has been investigated jointly with the beamformer and power allocations that maximize the \ac{EE} of the \ac{SBS}. Numerical results have shown that significant \ac{AEE} gain is attained by the system that allows \ac{SEg} harvesting.

\renewcommand{\baselinestretch}{1}
\bibliographystyle{IEEEtran}

\end{document}